\documentclass[useAMS,usenatbib,usegraphicx]{mn2e}
\usepackage{multirow}

\title[Microlensing of Central Lensed Images]
{Microlensing of Central Images in Strong Gravitational Lens Systems}
\author[Gregory Dobler, Charles R. Keeton, \& Joachim Wambsganss]
{Gregory Dobler$^{1,2}$, Charles R. Keeton$^3$, \& Joachim Wambsganss$^4$ \\
$^1$Department of Physics and Astronomy, University of Pennsylvania,
209 S.\ 33rd Street, Philadelphia, PA 19104 USA \\
$^2$Harvard-Smithsonian Center for Astrophysics, 60 Garden Street, MS-51,
Cambridge, MA 02138 USA \\
$^3$Department of Physics and Astronomy, Rutgers University,
136 Frelinghuysen Road, Piscataway, NJ 08854 USA \\
$^4$Astronomisches Rechen-Institut, Zentrum f\"ur Astronomie,
Universit\"at Heidelberg, M\"{o}nchhofstrasse 12-14, 69120 Heidelberg,
Germany}

\newcommand\reffig[1]{Figure \ref{fig:#1}}

\newcommand\Rein{R_{\rm E}}
\newcommand\Rsrc{R_{\rm src}}

\begin{document}

\maketitle

\begin{abstract}
We study microlensing of the faint images that form close to the
centers of strong gravitational lens galaxies.  These central images,
which have finally begun to yield to observations, naturally appear
in dense stellar fields and may be particularly sensitive to fine
granularity in the mass distribution.  The microlensing magnification
maps for overfocussed (i.e., demagnified) images differ strikingly
from those for magnified images.  In particular, the familiar ``fold'' and ``cusp''
features of maps for magnified images are only present for certain 
values of the fraction $f$ of the surface mass density contained in
stars.  For \emph{central} images, the dispersion in microlensing
magnifications is generally larger than for normal (minimum and
saddle) images, especially when the source is comparable to or
larger than the stellar Einstein radius.  The dispersion depends
in a complicated way on $f$; this behaviour may hold the key to
using microlensing as a probe of the relative densities of stars
and dark matter in the cores of distant galaxies.  Quantitatively,
we predict that the central image C in PMN J1632$-$0033 has a
magnification dispersion of 0.6 magnitudes for $\Rsrc/\Rein \la 1$,
or 0.3 mag for $\Rsrc/\Rein = 10$.  For comparison, the dispersions
are 0.5--0.6 mag for image B and 0.05--0.1 mag for image A, if
$\Rsrc/\Rein \la 1$; and just 0.1 mag for B and 0.008 mag for A
if $\Rsrc/\Rein = 10$.  (The dispersions can be extrapolated to
larger sources sizes as $\sigma \propto \Rsrc^{-1}$.)  Thus,
central images are more susceptible than other lensed images to
microlensing and hence good probes for measuring source sizes.
\end{abstract}


\section{Introduction}

Two aspects of strong gravitational lensing that date back 25 years
have attracted much recent interest.  Microlensing, or variations
in the (optical and/or x-ray) fluxes of lensed quasar images that
are induced by individual stars in the lens galaxy, was first
discussed by \citet{chang}.  The phenomenon has now been observed
in several lens systems
\citep{wozniak,s1104,richards,gaynullina05,kochanek06,morgan}.
It can be used to learn about the stellar components
\citep{gott,wps,pelt,schmidt,wyithe,csk}
and the relative densities of stars and dark matter \citep{SW,SW04}
in lens galaxies, and to probe structure in the source quasars at
stunning micro-arcsecond resolution
\citep{grieger,grieger2,agol,mineshige,fluke,richards,GSGO}.

Central or ``odd'' lensed images were originally predicted by
\citet{dyer} and \citet{burke}.  If the inner surface mass density
of a lens galaxy is shallower than $\Sigma \propto R^{-1}$ and there 
is no point mass at the center, then
lensing should always produce an odd number of images.  Nearly all
observed lenses, however, exhibit two or four images.  The prediction
and observations can be reconciled by noting that one of the
expected images should be very close to the center of the lens
galaxy and demagnified by the high central surface density there
\citep[e.g.,][]{wallington,norbury,rusinma,centers}.  That makes
central images difficult to detect --- and indeed the first ones
have only just been found.  \citet{winn03,winn04} detected a
central image at radio wavelengths in the asymmetric 2-image lens
PMN J1632$-$0033, while \citet{inada} detected a central image at
optical wavelengths in the unusual 4-image cluster lens
SDSS J1004+4112.

Our goal in this paper is to combine the two phenomena and study
microlensing of central images.  The issue is timely
since central images are now being observed, and compelling since
the images naturally form in regions where the density of stars is
high and microlensing seems inevitable.

We customize our calculations to the lens PMN J1632$-$0033, because
it is both a known central image system and also a prototype for the
sorts of systems that are expected to yield the most central images
in the future \citep{centers,bowman}.  \citet{winn03,winn04} recently
studied a wide range of mass models for PMN J1632$-$0033, which can
be used to estimate the total convergence $\kappa$ and shear $\gamma$
at the positions of the images.  The lens data, including the central
image, are consistent with a simple power law surface mass density
$\Sigma \propto R^{-\alpha}$ with $\alpha = 0.91 \pm 0.02$, plus a
small external tidal shear.  Table \ref{tbl:kapgam} gives the values
of $\kappa$ and $\gamma$ for such a model, which we adopt as inputs
for our microlensing simulations.

\begin{table}
\begin{center}
  \begin{tabular}{crrr}
\hline
\multicolumn{1}{c}{Image} &
\multicolumn{1}{c}{$\kappa$} &
\multicolumn{1}{c}{$\gamma$} &
\multicolumn{1}{c}{$\mu$} \\
\hline
\hline
 A &  0.370 &  0.422 & 4.570 \\
 B &  3.309 &  2.859 & 0.352 \\
 C & 16.517 & 13.796 & 0.020 \\
\hline
  \end{tabular}
\caption{
Total convergence $\kappa$, shear $\gamma$, and magnification $\mu$
for each of the three images in PMN J1632$-$0033, from spherical 
power law plus external shear lens models by \citet{winn03,winn04}.  
``C'' indicates the central, highly demagnified image.
}\label{tbl:kapgam}
\end{center}
\end{table}

\section{Methods}

We use standard methods for microlensing calculations \citep{jcam}.
Specifically, we pick a patch around an image that is large compared
with a stellar Einstein radius (the scale for microlensing), but
small compared with the global scale of the lens so that the mean
densities of stars and dark matter are essentially constant across
the patch.  Attaining this balance is not difficult because the
global scale is set by the $1\farcs5$ image separation while the
stellar Einstein radius is
$\Rein \sim 2\,(M/M_\odot)^{1/2} \times 10^{-6}$ arcsec.  In
practice, we typically consider a rectangular patch in the image
plane chosen so that it maps into a square patch in the source
plane that is $100 \Rein$ on a side (see below).  Without loss of
generality, we can choose a coordinate system aligned with the
direction of the local shear.  We assume that a fraction
\begin{equation}
  f = \kappa_{\rm stars}/\kappa_{\rm tot}
\end{equation}
of the total surface mass density (or convergence) is contributed
by stars.  We use that to determine the number density of stars in
the image plane patch,
\begin{equation}
  n_* = \frac{f\,\kappa_{\rm tot}}{\pi \Rein^2}\ ,
\end{equation}
and then distribute the stars randomly.  The total surface
density $\kappa_{\rm tot}$ is kept fixed (to the values in
Table \ref{tbl:kapgam}) by including a smooth, continuous
matter component with $\kappa_{c} = (1-f) \kappa_{\rm tot}$.
\citet{pac} has shown that including a smooth matter component is equivalent to rescaling the convergence and shear 
to
\begin{equation}
  \left( \kappa^{\rm eff}, \gamma^{\rm eff} \right) = \frac{\left(\kappa, \gamma \right)}{\left| 1-\kappa_{c} \right|},
\end{equation}
with $f = 100\%$ in stars.  For the case of a finite source, this also leads to a rescaling of the source size,
\begin{equation}
  \Rsrc^{\rm eff} = \frac{\Rsrc}{\left| 1-\kappa_{c} \right|}.
\end{equation}

In this pilot study, we assume that all stars have the same mass,
and we always work in units of the stellar Einstein radius $\Rein$.
Previous analyses have shown that the microlensing magnification
distribution is at best weakly dependent on the distribution of
stellar masses, at least for a point source.\footnote{The
magnification distribution was long thought to be independent of
the microlens mass function \citep{s87,w92,LI,WT}.  However,
\citet{SWL} recently presented results from numerical experiments
suggesting that there is a weak dependence.}  That result must break
down for a finite source, which would be insensitive to stars below
some mass threshold (roughly corresponding to $\Rein \la \Rsrc$).
The full problem --- microlensing of a finite source by stars with
unequal masses --- is certainly interesting, but beyond the scope
of this paper \citep[cf.][]{congdon06}.

We use ray-shooting software by \citet{w90}, \citet{wps}, and
\citet{jcam} to perform the microlensing simulations.  Briefly,
the software ``shoots'' a large number ($\ga\!10^{8}$) of light
rays from the observer through the lens/image plane into the
source plane, and collects them in small pixels.  The number of
rays in a source pixel is proportional to the lensing magnification
at that position.  As described by \citet{jcam}, the rectangular
shooting patch in the image plane is larger than the region that
would map into the source plane box in the absence of microlensing.
The reason is that the deflection diverges as $1/r$ near a star,
so stars outside the nominal shooting region can send light rays
into the source plane box \citep{kbp}.  The shooting region is expanded to
ensure that a significant fraction of this ``diffuse flux'' is
collected.

The ray-shooting software produces a magnification map in the
source plane.  We can obtain the magnification probability
distribution by making a histogram of the pixel magnifications.
To consider a source with a finite extent, we convolve the
magnification map with the surface brightness distribution of
the source before making the histogram.  For simplicity, we use
a Gaussian source with half-light radius $\Rsrc$, since
\citet{mortonson} argue that the detailed structure of the source
hardly affects magnification distributions provided that the
half-light radius is the same \citep[also see][]{congdon06}.

The physical input parameters for the simulations are the total
convergence and shear, which we take from Table \ref{tbl:kapgam},
and the fraction $f$ of mass in stars, which we vary (see \S 3).
The technical input parameters are the size of the magnification
map and the number of pixels.  We seek to consider sources both
smaller and larger than $\Rein$, and in particular find the range
$0.1 \le \Rsrc/\Rein \le 10$ to be both interesting and tractable.
In order to have many independent source positions in a given
magnification map, we use maps that are $100 \Rein$ on a side.
In order to have enough pixels to handle small sources, we use
maps with $1024 \times 1024$ pixels.

To obtain fair sampling, we create 50 independent magnification maps
for each image, and compute the final magnification distributions
from the combination of all of them.  A fringe benefit of this
approach is that we can consider the 50 magnification maps to be
independent realizations of the microlensing calculation, and use
bootstrap or jackknife resampling \citep[e.g.,][]{efron} to estimate
the statistical uncertainties in our analysis.

\section{Results}

\subsection{Magnification maps}
\label{sec:maptop}

\begin{figure*}
\begin{center}
\includegraphics[width=0.9\textwidth]{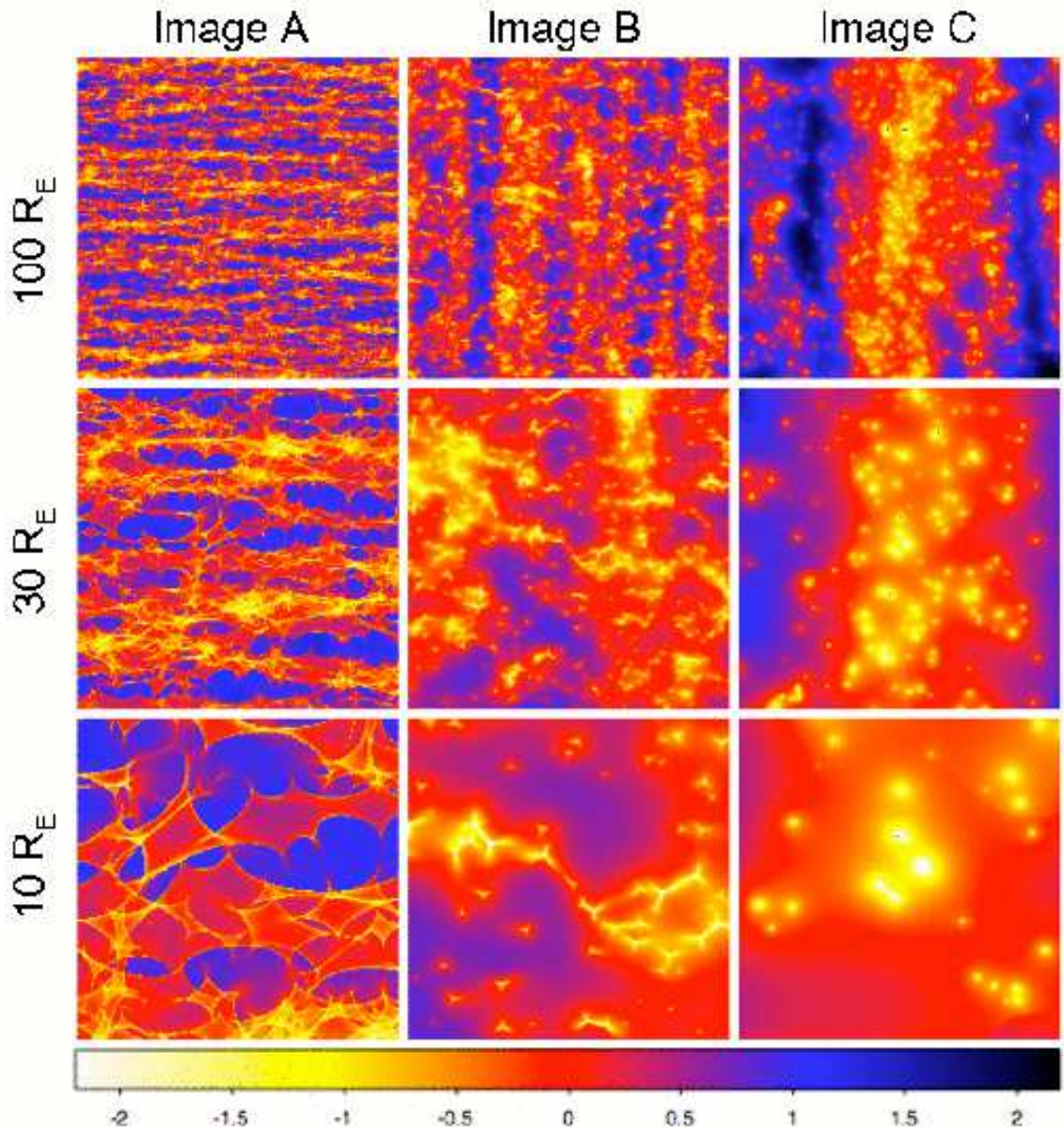}
\end{center}
\caption{
Sample magnification maps for images A (left), B (center), and C (right).
The top row shows $100 \Rein$ on a side, the middle row shows $30 \Rein$ on 
a side, and the bottom row shows $10 \Rein$ on a side.  The zoom in region in each case is centered on the $(100 \Rein)^2$ 
map.  The fraction of 
surface mass
density in stars is $f=100\%$ for all cases, and the colorbar at the bottom indicates
the change in magnitude, with respect to the mean magnification, at each 
source position (yellow regions indicate magnification, blue regions indicate demagnification).
}\label{fig:magmaps}
\end{figure*}

\reffig{magmaps} shows magnification maps for $f=100\%$ of the
surface density in stars.
To display the range of features that appear on different length scales,
we show $(100 \Rein)^2$, $(30 \Rein)^2$, and $(10 \Rein)^2$ maps here.  The 
maps for image C in particular show that there is structure on
scales of tens of Einstein radii, so even a $(100 \Rein)^2$ map
does not contain a fully representative sample of magnifications.

The qualitative differences in the three magnification maps shown
in \reffig{magmaps} are striking.  For image A, which forms at a
local minimum in the time delay surface and has relatively low
values of the convergence and shear leading to a modest
amplification, the map shows the familiar caustic network
\citep[e.g.,][]{w90,SEF,jcam}.  The caustics are preferentially
stretched along the horizontal axis because we have chosen
coordinates aligned with the local shear.  For image B, which
forms at a saddle point in the time delay surface and is modestly
demagnified, careful inspection reveals that the map shows many of
the 3-pointed cusps that are characteristic of microlensing when
the parity is negative \citep[e.g.,][]{SEF,PLW}.  These results
are familiar.  More surprising is the map for image C, which forms
at a local maximum in the time delay surface and is highly demagnified.
This map looks very different from the familiar, traditional caustics.
It exhibits round, concentrated blobs of high magnification,
surrounded by large regions of low magnification.  These features
are known to occur for highly overfocussed (i.e., demagnified)
microlensing scenarios \citep[e.g.,][]{w90,SEF,petters96,PLW}, but
they have not been explored in great
detail.\footnote{\citet{petters96} and \citet[see Chapter 15]{PLW}
have shown that, in the case of an isolated star with a constant
shear and a large, constant convergence $\kappa_{c}$, the caustics
consist entirely of elliptical fold curves and do not have any cusps.}

\begin{figure*}
\begin{center}
\includegraphics[width=0.8\textwidth]{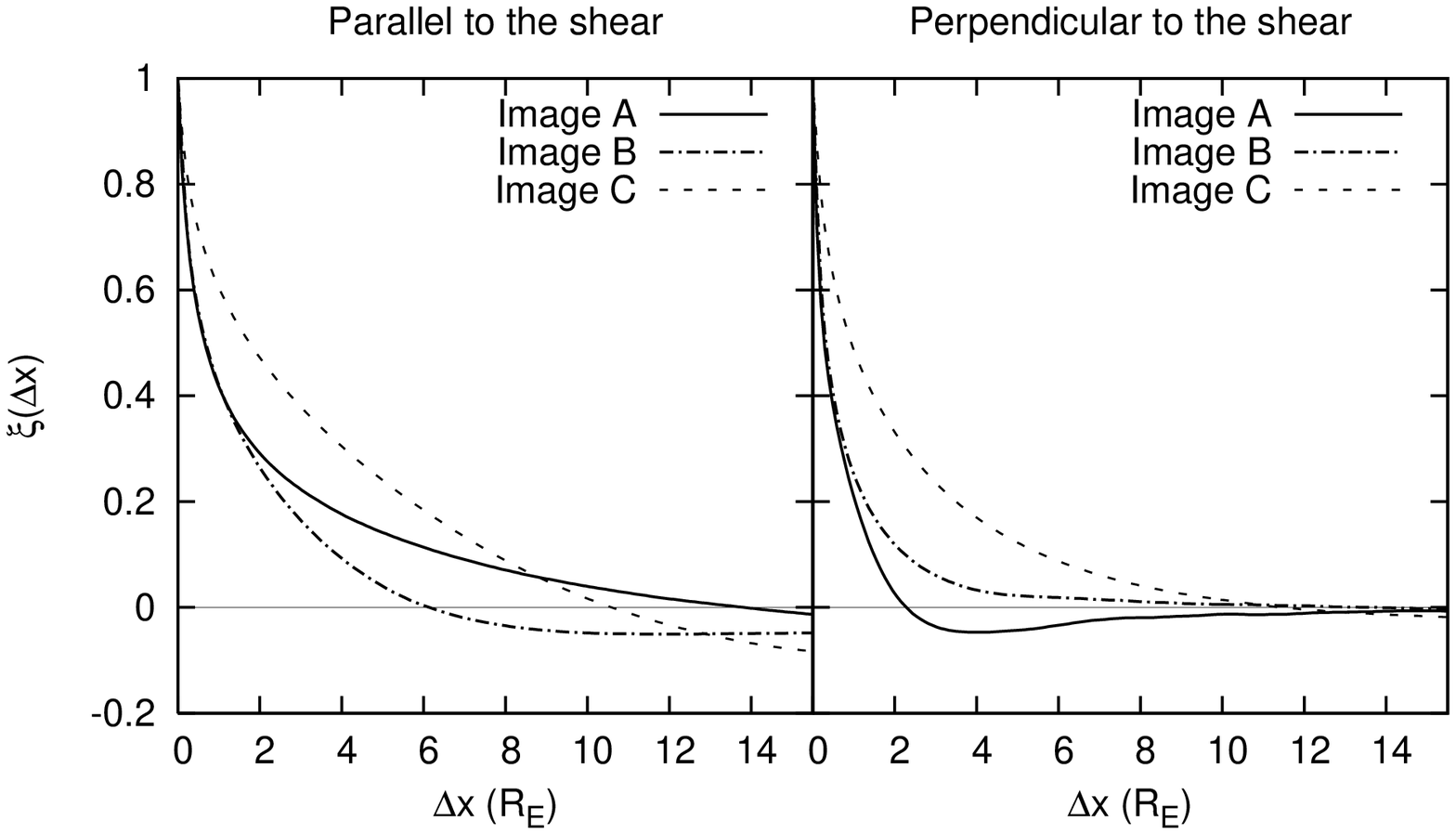}
\end{center}
\caption{
One-dimensional auto-correlation functions for magnification maps
with $f=100\%$.  The left panel shows $\xi(\Delta x)$ aligned with
the shear axis, while the right panel shows $\xi(\Delta x)$
perpendicular to the shear axis.  In the left panel, there is
significant power in $\xi^{\rm A}$ because the caustics tend to be
stretched along the shear axis.  The fact that
$\xi^{\rm C} > \xi^{\rm B}$ implies that there is structure in the
image C maps on larger length scales compared to the image B maps.
}\label{fig:corrfunc} 
\end{figure*}

As noted above, there appears to be strong inhomogeneity even on
scales of tens of Einstein radii.  To quantify this, we define the
auto-correlation function along linear slices through the maps
\citep[see][]{WPK}:
\begin{equation}
  \xi(\Delta x) = \frac{\langle \mu(x)\mu(x+\Delta x) \rangle
    - \langle \mu(x) \rangle^2}
    {\langle \mu(x)^2 \rangle - \langle \mu(x) \rangle^2}\ ,
\end{equation}
where $\mu(x)$ is the magnification at position $x$ and the averages 
are over all positions along the slice.  For each map, we calculate
$\xi(\Delta x)$ for each row and average over rows.  We then further
average over 50 maps (i.e., 50 random stellar configurations).  This
gives $\xi(\Delta x)$ parallel to the shear axis; we then repeat the
procedure averaging over columns to find $\xi(\Delta x)$ perpendicular
to the shear.  The auto-correlation function for each image, both
parallel and perpendicular to the shear, is shown in
Figure \ref{fig:corrfunc}.  The distance $\Delta x_{0}$ at which
$\xi(\Delta x)$ falls to zero is a measure of the correlation
length.

Parallel to the shear, $\Delta x_{0}^{\rm B} \approx 6.1 \Rein$ 
while $\Delta x_{0}^{\rm C} \approx 10.6 \Rein$, implying that the
image B maps have structures spanning $\sim$10--15 $\Rein$ and the
image C maps have structures $\sim$20--25 $\Rein$.  It is difficult
to interpret $\xi^{\rm A}$ parallel to the shear since the caustics
are preferentially stretched along the shear axis, leading to
substantial power in $\xi^{\rm A}$ even at large $\Delta x$.
Perpendicular to the shear, $\xi^{\rm A}$ falls rapidly to zero 
($\Delta x_{0}^{\rm A} \approx 2.3 \Rein$ along this direction) 
and in fact Figure \ref{fig:magmaps} shows little in the way of
large scale structure in the vertical direction.  For images C and
B, we see that $\xi^{\rm C} > \xi^{\rm B}$ for all distances $\Delta x$.  
This supports the qualitative conclusion that there is more structure
on large scales for the image C compared to image B maps.

\begin{figure*}
\begin{center}
\includegraphics[width=0.9\textwidth]{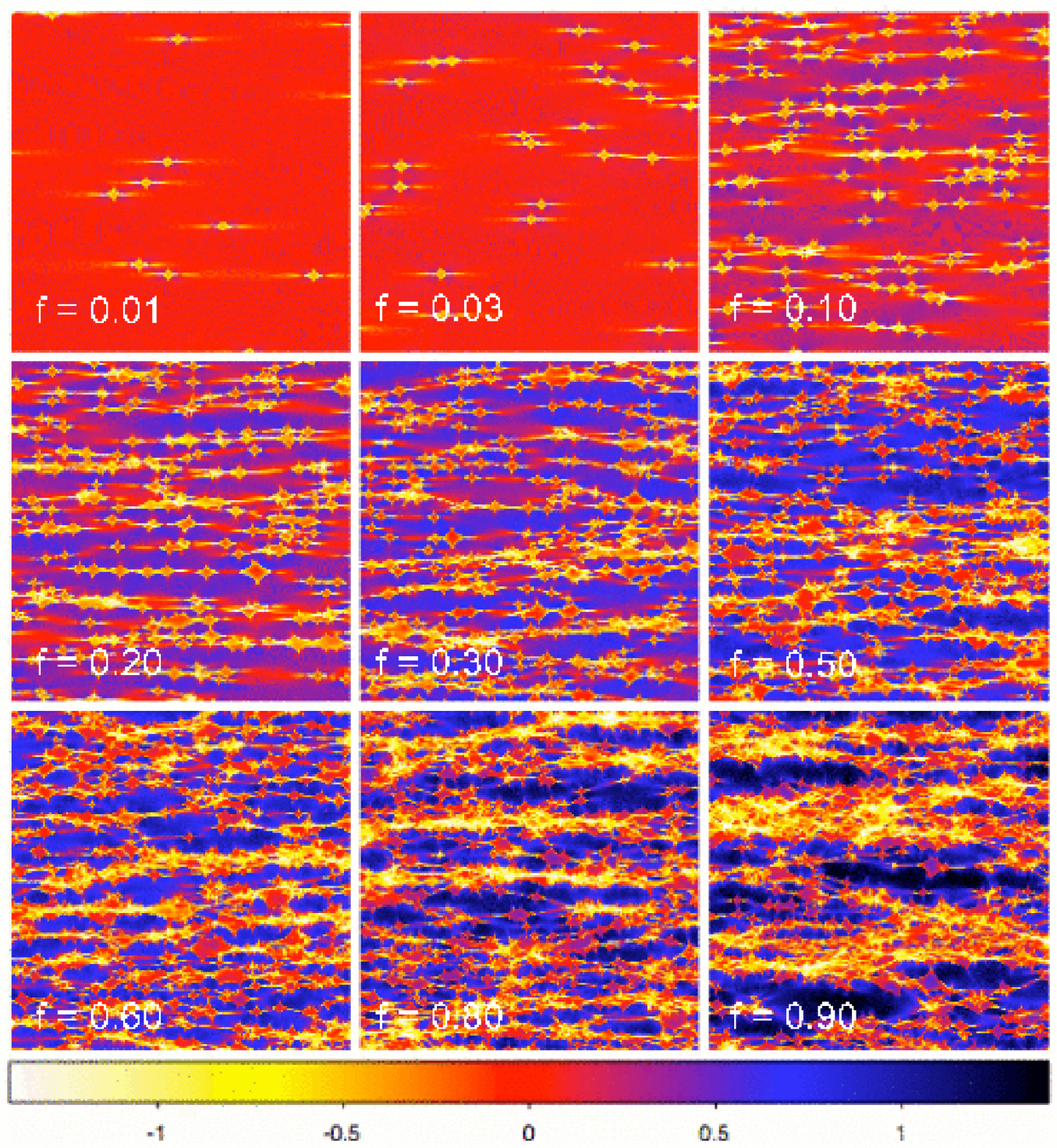}
\end{center}
\caption{
Magnification maps for image A with size $(50 \Rein)^2$ and varying $f$ 
(cf. Figure \ref{fig:magmaps}, left column, for $f = 1.0$)
The visual contrast in each
map conveys the magnification dispersion.  The dispersion increases
with increasing $f$: rapidly for small $f$, more slowly above $f \ga 0.3$
(compare to \reffig{sigvf} with $\Rsrc=0.1$).
}\label{fig:dispmapsA}
\end{figure*}

\begin{figure*}
\begin{center}
\includegraphics[width=0.9\textwidth]{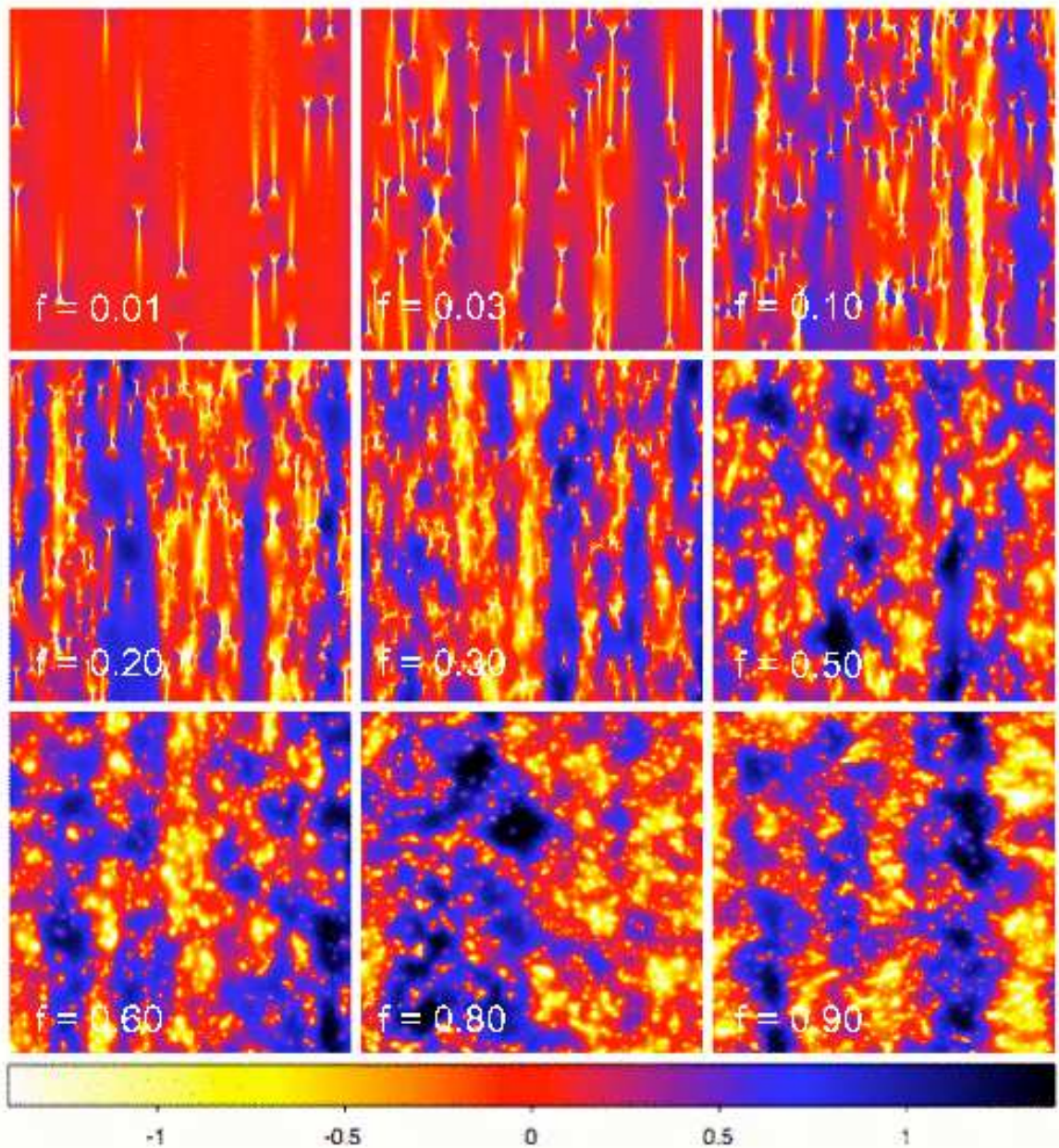}
\end{center}
\caption{
The same as \reffig{dispmapsA}, but for image B
(cf. Figure \ref{fig:magmaps}, middle column, for $f = 1.0$).
The relative number of cusps and folds
versus blob caustics has a complicated dependence on $f$.  The visual
contrast in each map conveys the magnification dispersion.  The dispersion
increases 
with increasing $f$: rapidly for small $f$, more slowly above $f \ga 0.3$
(compare to \reffig{sigvf} with $\Rsrc=0.1$).
}\label{fig:dispmapsB}
\end{figure*}

\begin{figure*}
\begin{center}
\includegraphics[width=0.9\textwidth]{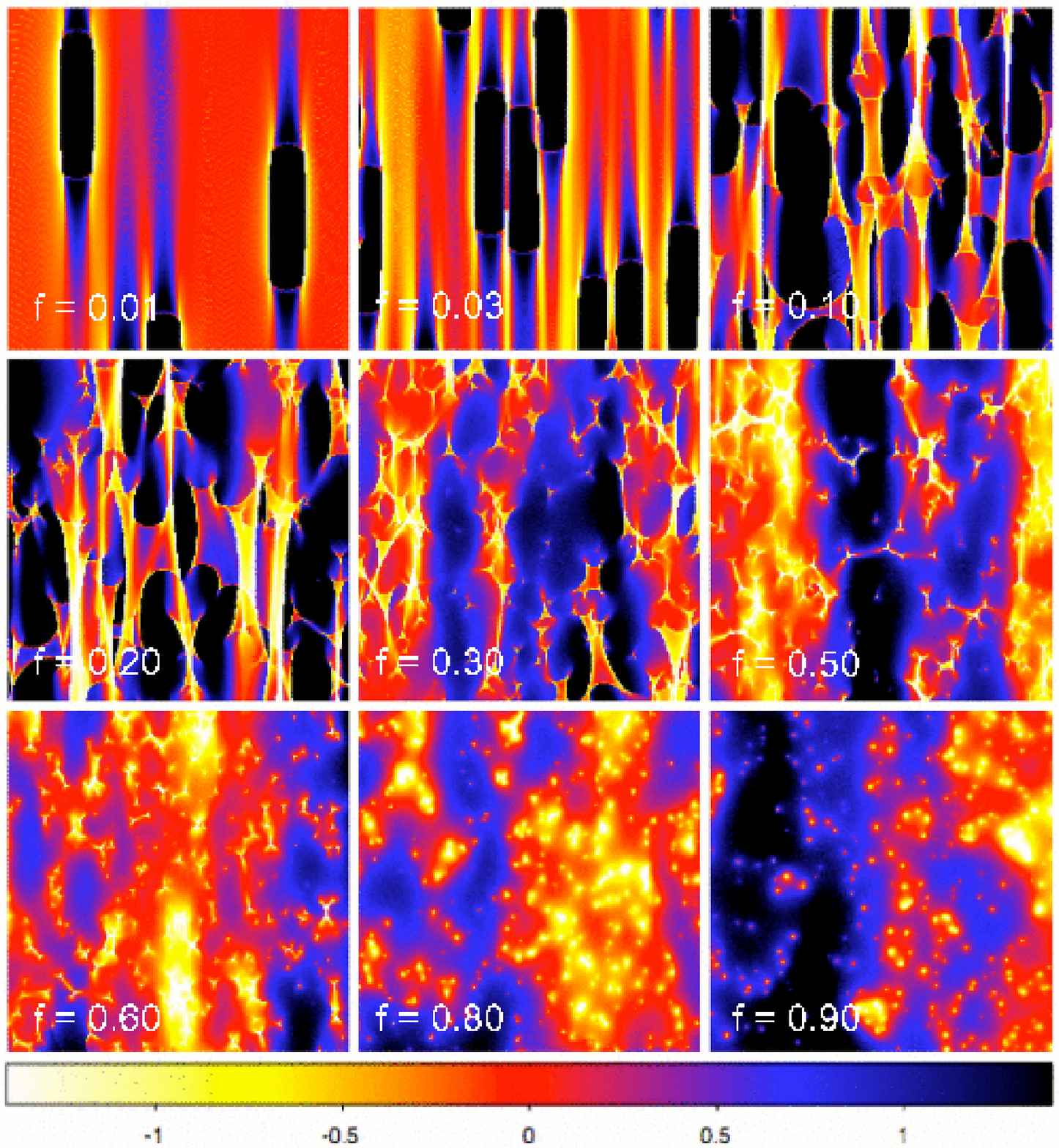}
\end{center}
\caption{
The same as \reffig{dispmapsA}, but for image C
(cf. Figure \ref{fig:magmaps}, right column, for $f = 1.0$).
The relative number of cusps and
folds versus blob caustics has a complicated dependence on $f$.  The visual
contrast in each map conveys the magnification dispersion.  In contrast to
Figures \ref{fig:dispmapsA} and \ref{fig:dispmapsB}, the dispersion
\emph{decreases} with $f$, rapidly at first and then more slowly above
$f \ga 0.4$ (compare to \reffig{sigvf} with $\Rsrc=0.1$).
}\label{fig:dispmapsC}
\end{figure*}

So far we have assumed that $f=100\%$ of the surface mass density
for each image is in stars, which is probably valid for image C but
unlikely for image A.  In Figures \ref{fig:dispmapsA}--\ref{fig:dispmapsC}
we examine the magnification maps for different values of $f$.
In \reffig{dispmapsA}, when $f$ is low, the map exhibits caustics of
individual stars, and as $f$ increases the number of caustics
increases and they begin to merge \citep[cf.][]{SW}.

Figures \ref{fig:dispmapsB} and \ref{fig:dispmapsC} show drastically different behaviour.
For image C (Figure \ref{fig:dispmapsC}), the caustics morph from blobs to cusps as $f$ decreases; by
$f=0.2$ the magnification map consists entirely of fold and cusp
caustics.  Decreasing $f$ further causes holes to appear in the
magnification map \citep{chang2}.  For image B, we find the curious result that cusp
caustics are present for $f=1.0$ (Figure \ref{fig:magmaps}), but they are absent for
$0.9 \ga f \ga 0.5$, and then they reappear at lower $f$
values.  This behaviour is surprising because it was not seen by \citet{SW} in their magnification maps for a
highly magnified saddle image.  Since our image B is a demagnified
saddle image, we conjecture that the overall magnification, in
addition to the parity, is important for determining whether the
caustics are blobs or cusps and folds.  The dependence on the stellar
mass fraction appears to be complicated.

\subsection{Magnification distributions}

\begin{figure*}
\begin{center}
\includegraphics[width=0.85\textwidth]{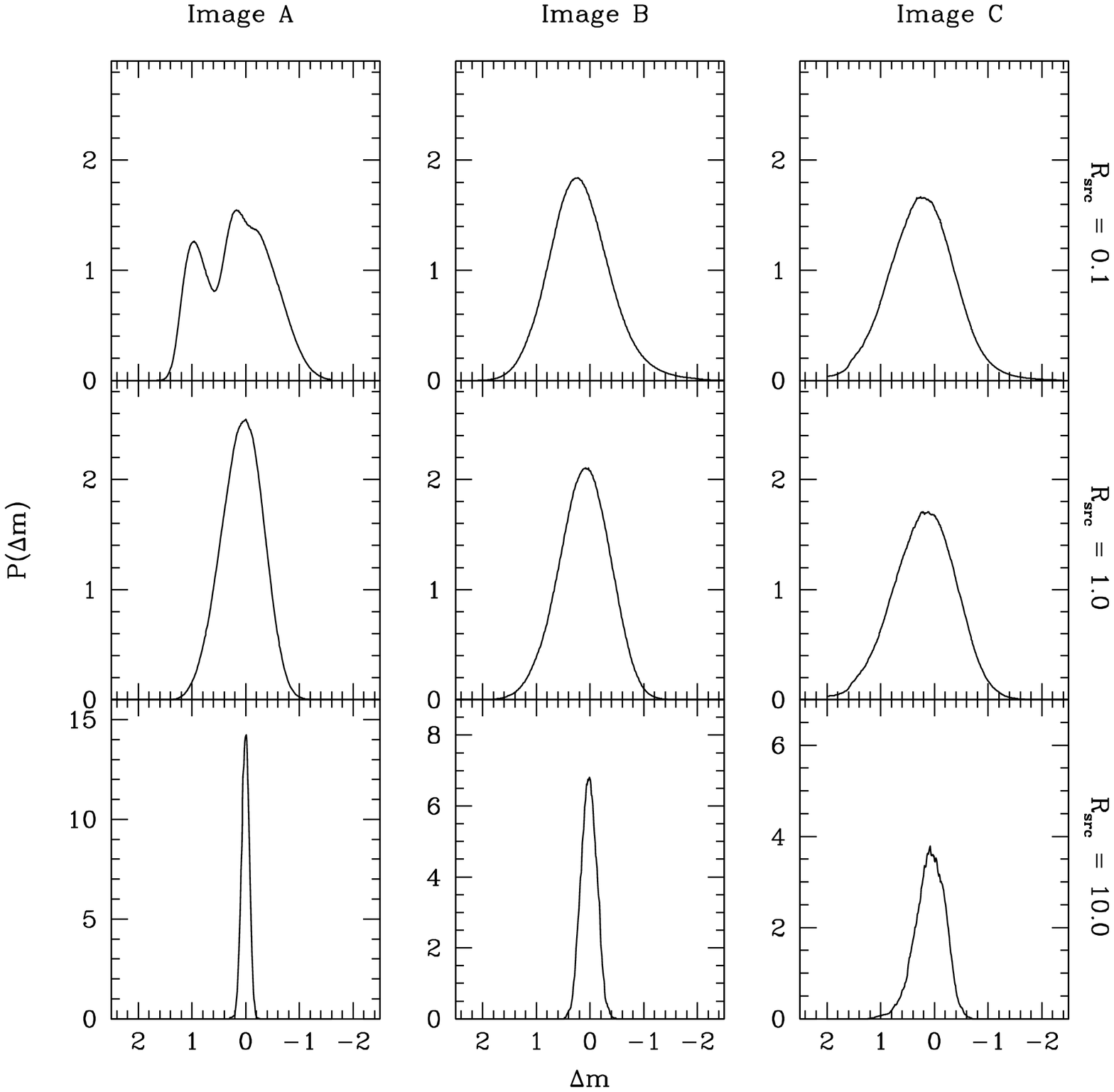}
\end{center}
\caption{
Magnification distributions for the three images (left to right),
for three different source sizes (top to bottom), assuming $f=100\%$
of the mass in stars.  The source is modeled as a Gaussian with
half-light radius $\Rsrc$ (quoted in units of $\Rein$).  The
distributions for each image are derived from 50 independent
realizations of $(100\Rein)^2$ magnification maps.  Each
probability distribution $P(\log\mu)$ is normalized to unit area,
but the vertical axis scale varies in the bottom panels.  Any
apparent jaggedness is numerical noise.  
The double peak in the panel at the top left is real
and reflects the first additional microimage pair (for details see text).
The histograms are centered on 
$\Delta m = -2.5\log(\mu/\mu_0) = 0$, where $\mu_0$ is the magnification in the absence
of microlensing (from Table \ref{tbl:kapgam}.)  
The microlensing distributions yield a mean magnification $\langle\mu\rangle$ that
agrees with $\mu_0$ to within numerical uncertainties.
}\label{fig:histograms}
\end{figure*}

The magnification distributions for different source sizes and
different images are shown in \reffig{histograms} (assuming
$f=100\%$).  The magnification is expressed as the change in 
magnitude due to the presence of microlensing, $\Delta m = -2.5\log(\mu/\mu_0)$.
In general, the distributions appear to be smooth
and fairly symmetric about $\Delta m = 0$.  The exception is the case of
image A with a small source ($\Rsrc/\Rein \la 0.1$), whose
magnification distribution shows three distinct peaks.  The peaks
can be understood in terms of the caustic network
\citep[e.g.,][]{rauch,wws,granot}:  from high to low $\Delta m$ --- i.e., low to high $\mu$ ---, the first
peak corresponds to sources outside all caustics, the second peak
to sources inside a single caustic (so there is one additional
pair of microimages), and the third peak to sources inside a pair
of overlapping caustics (so there are two extra microimage pairs).
In fact, there are higher-order peaks corresponding to source
positions where more caustics overlap, but those peaks blur
together and do not appear as clear features in the final
magnification distribution
\citep[see][especially their Fig.\ 4]{granot}.  The multi-peak
structure disappears as the source size increases, because a large
source ($\Rsrc/\Rein \ga 1$) generally extends over one or more
caustics, so the regions of different image multiplicity in the
source plane are smeared out and disappear as distinct maxima in
the magnification distribution.  The multi-peak structure is not
apparent in the histograms for images B and C, even for sources
as small as $\Rsrc = 0.1\Rein$, because the caustics are so small
that only a small fraction of source positions lead to additional
microimage pairs.

An important qualitative feature of the magnification distributions
is that the width decreases as the source size increases.  The
width can be considered to represent either the likelihood that
the magnification at any given time is different from what a
smooth model would predict, or the RMS amplitude of variations in
time.  Thus, we may say that the larger the source, the less it is
affected by microlensing --- a well known result which makes intuitive sense.  To
quantify this effect, we compute the dispersion in magnitudes,
namely
\begin{equation}
  \sigma = \left[\langle(-2.5\log\mu)^2\rangle -
  \langle -2.5\log\mu\rangle^2 \right]^{1/2} ,
\end{equation}
where the average is over source positions.  In previous work
the magnification dispersion has been studied analytically
\citep[for the case of $\gamma = 0$, $\kappa_c \la 1$, and
$\Rsrc \ga 5\Rein$;][]{RS1,sws,nein},
numerically \citep{DW,RS1,sws,RS2,WT2}, and with ray-shooting
methods \citep{SW,WT2,lew-iba,SWL,mortonson}.

\begin{figure}
\begin{center} 
\includegraphics[width=0.4\textwidth]{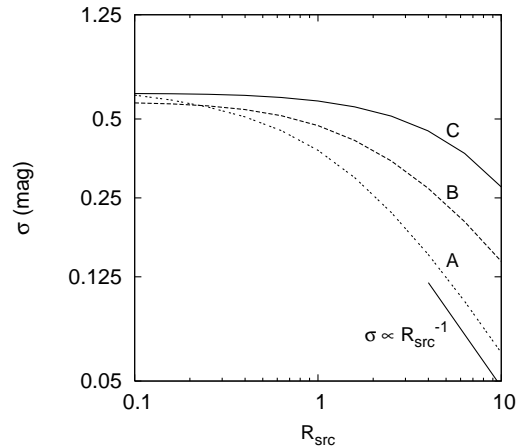}
\end{center} 
\caption{
Magnification dispersion as a function of source size, for the
three images, assuming $f=100\%$ of the mass in stars.  The line
segment below the curve for image A shows the expected asymptotic
scaling for large sources, $\sigma \propto \Rsrc^{-1}$ \citep{RS1,RS2}.
}\label{fig:sigvR}
\end{figure}

\reffig{sigvR} shows that the dispersion falls monotonically as
the source size increases (as seen in the previous work).  It
remains relatively constant until the source size becomes comparable
to the stellar Einstein radius, and then drops rapidly.  Curiously,
all three images have roughly the same magnification dispersion when
the source is small ($\Rsrc/\Rein \la 0.3$).  For
$\Rsrc/\Rein \ga 0.3$, the ordering of the images is independent
of source size: image C is most affected by microlensing, followed
by image B, and finally image A.  For large sources analytic
arguments suggest that the dispersion should scale as
$\sigma \propto \Rsrc^{-1}$ \citep{RS1,RS2}.  We certainly see
this for image A.  It appears that images B and C are approaching
the asymptotic scaling but have not quite reached it at
$\Rsrc/\Rein = 10$.  This is not surprising because the predicted
scaling applies when the number of stars in front of the image is
large.  Since images B and C are demagnified, the source needs to be
larger (compared with the magnified image A) in order for the image
to be large enough to intercept a significant number of stars.

\begin{figure*}
\begin{center}
\includegraphics[width=0.32\textwidth]{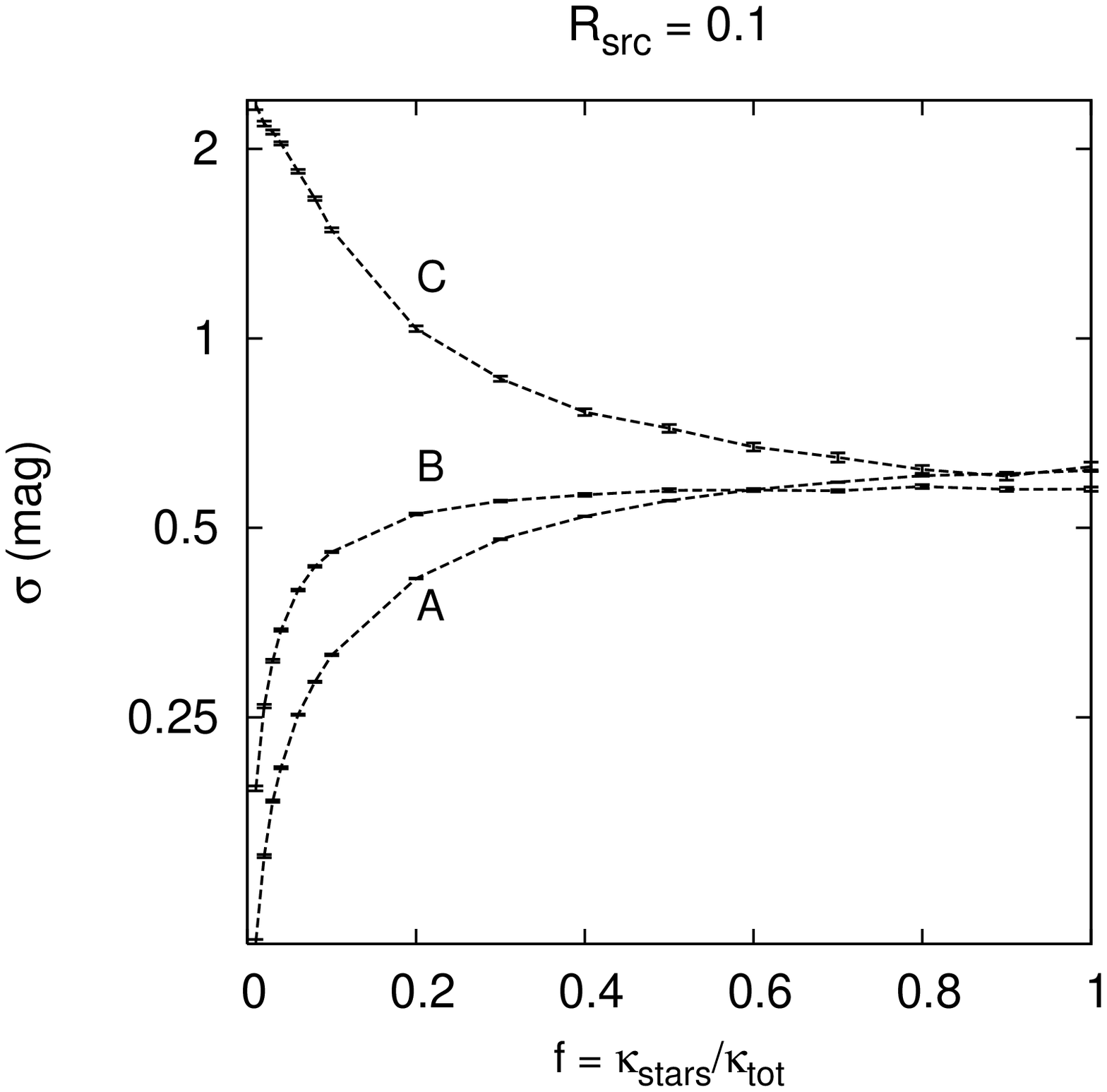}
\includegraphics[width=0.32\textwidth]{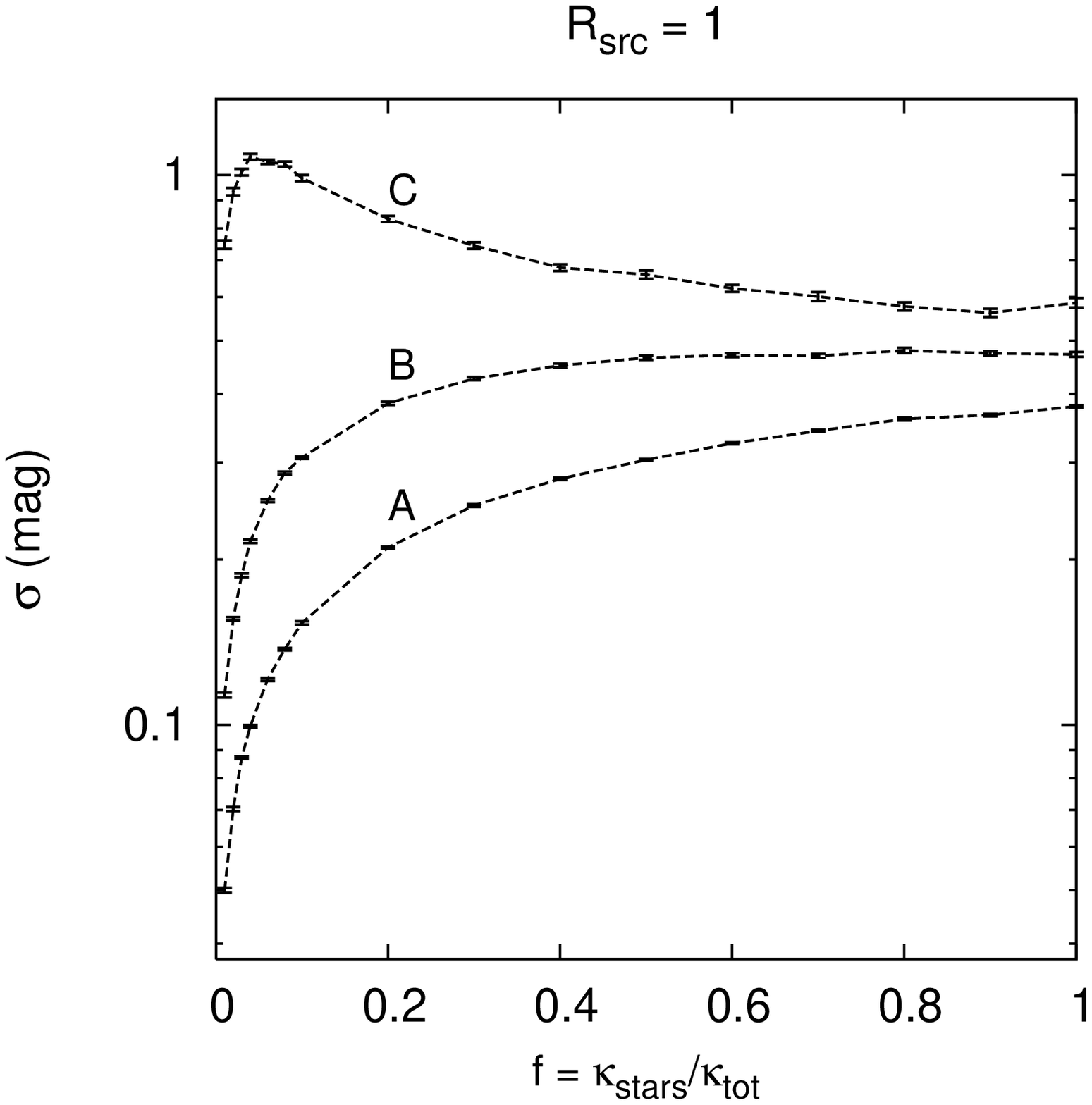}
\includegraphics[width=0.32\textwidth]{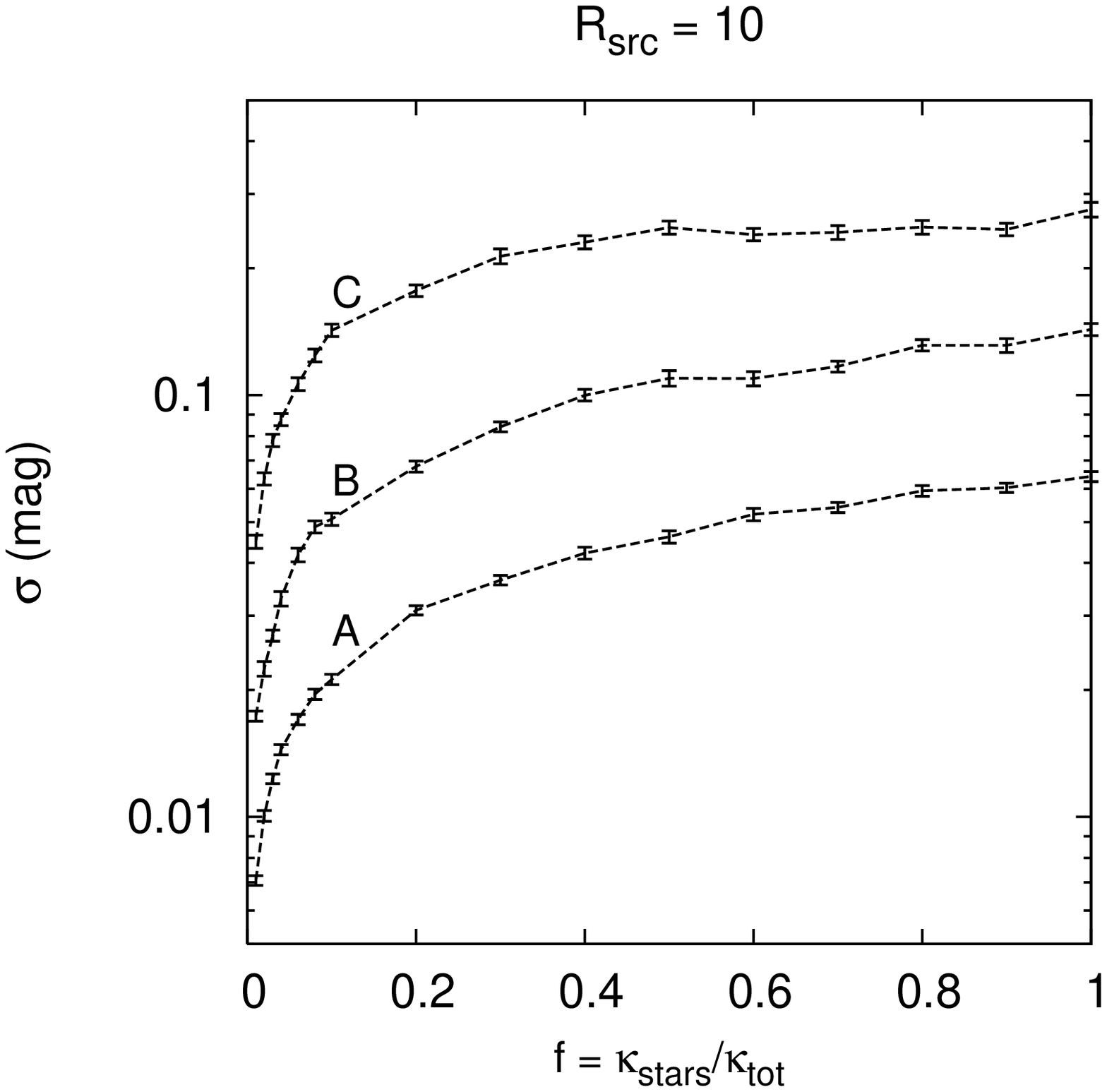}
\end{center}
\caption{
Magnification dispersion as a function of the fractional surface
density in stars, $f = \kappa_{\rm stars}/\kappa_{\rm tot}$, for
three different source sizes (quoted in units of $\Rein$).  
The three lines show the three images A (bottom),B (middle), and C (top) of PMN J1632$-$0033.
Note the widely different scales for $\sigma$.
The errorbars indicate the statistical uncertainties estimated with
bootstrap and jackknife resampling \citep[e.g.,][]{efron}.
}\label{fig:sigvf}
\end{figure*}

The next step is to consider how the magnification dispersion
changes as we decrease $f$, as shown in \reffig{sigvf}.  For
images A and B, the dispersion falls monotonically as $f$
decreases.  For image B, this is in contrast to \citet{SW} who
found that decreasing $f$ can \emph{increase} the magnification
dispersion for saddle images.  We believe the difference is
explained by the fact that \citeauthor{SW} considered a highly
magnified saddle image ($\mu = 9.5$), such as would be seen in
a bright minimum/saddle pair straddling a critical curve in a
4-image lens; whereas our saddle image B has a modest
demagnification ($\mu = 0.35$), as is common for the image
nearer the lens galaxy in an asymmetric 2-image lens.

For the central image C, we find that decreasing the fractional surface 
density in stars can \emph{increase} the magnification dispersion, so
long as $\Rsrc \la \Rein$.  Specifically, for a large source
$\Rsrc/\Rein = 10$ (right panel of \reffig{sigvf}), the magnification
dispersion for image C decreases monotonically as $f$ decreases, like
it does for images A and B.  When $\Rsrc/\Rein = 1$, however, the
dispersion increases as $f$ decreases all the way down to
$f \approx 0.04$, and only then turns over.  (The turnover is
reassuring, because for $f \to 0$ there is no microlensing and the
magnification dispersion must vanish.)  When $\Rsrc/\Rein = 0.1$,
the turnover occurs at $f < 0.01$.  While the behaviour of image C at
low $f$ may have little practical importance, it is still useful for
understanding how the magnification dispersion depends on the
relative densities of stars and dark matter.

We note that, in the limit of an infinitesimal source, the dispersion
is visually apparent in the contrast of the magnification maps shown
in Figures \ref{fig:dispmapsA}, \ref{fig:dispmapsB}, and \ref{fig:dispmapsC}.
The dispersion in the maps of images A and B visibly increases with
$f$, while the dispersion of the image C maps \emph{decreases} with
increasing $f$.  Further, the contrast changes sharply at low $f$ for
all three images and then remains roughly constant for $f \ga 0.4$.
This is consistent with the slopes of the $\sigma$ versus $f$ curves
shown in \reffig{sigvf}.

To make a final interpretation of \reffig{sigvf}, it is useful to
estimate an upper bound on the fractional surface density in stars
at the position of each image.  For PMN J1632$-$0033, \citet{winn03}
show that the lens data are consistent with a total surface mass
density of the form $\kappa_{\rm tot} \propto R^{-\alpha}$ with
$\alpha = 0.91 \pm 0.02$.  They note that the lens galaxy is too
faint to be characterized in detail with existing HST images, but
appears to be an early-type galaxy with effective radius of
$R_e \approx 0\farcs2$.  To estimate the maximum possible stellar
mass density as a function of radius, we model $\kappa_{\rm stars}$
using a de Vaucouleurs $R^{1/4}$ law profile, and set the stellar
mass-to-light ratio to the largest allowed value such that
$\kappa_{\rm stars}$ never exceeds $\kappa_{\rm tot}$.  Using the
observed distances to the center of the lens ($R_A = 1\farcs38$,
$R_B = 0\farcs09$, and $R_C = 0\farcs01$; see \citealt{winn03}),
we obtain the $f_{\rm max}$ values listed in Table \ref{tbl:smax}.

\begin{table}
\begin{center}
  \begin{tabular}{cccll}
\hline
\multirow{2}{*}{Image} &
\multirow{2}{*}{$f_{\rm max}$} &
\multicolumn{3}{c}{$\sigma$ (in mag)} \\
& &
\multicolumn{1}{c}{$\Rsrc = 0.1$} &
\multicolumn{1}{c}{1} &
\multicolumn{1}{c}{10} \\
\hline
\hline
 A & $0.013 \pm 0.001$ & 0.13 & 0.05 & 0.008 \\
 B & $0.511 \pm 0.023$ & 0.58 & 0.48 & 0.10  \\
 C & $0.997 \pm 0.003$ & 0.63 & 0.58 & 0.28  \\
\hline
  \end{tabular}
\caption{
Column 2 gives our estimate of the upper limit on
$f = \kappa_{\rm stars}/\kappa_{\rm tot}$ (see text).  The quoted 
uncertainties are derived from the uncertainty in the power law index 
of the lens galaxy mass distribution 
\citep[$\alpha = 0.91 \pm 0.02$;][]{winn03}.  Columns 3--5 
give upper limits on the magnification dispersion $\sigma$ (in magnitudes),
for three source sizes (quoted in units of $\Rein$).  These upper
limits were obtained by combining $f_{\rm max}$ with the $\sigma$
vs.\ $f$ curves in \reffig{sigvf}.  The uncertainties in $\sigma$ due 
to the uncertainties in $f_{\rm max}$ are $\ll 0.13$ mag.  The 
dispersion can be extrapolated to larger source sizes with the scaling 
$\sigma \propto \Rsrc^{-1}$ \citep{RS1,RS2}.
}\label{tbl:smax}
\end{center}
\end{table}

This analysis confirms our guess that the matter at image C could
be essentially all stars.  By contrast, the galaxy appears to be
concentrated enough that at image B the density is no more than
half stars, and at the distance corresponding to image A the density
is no more than 1\% stars.  When we combine the upper limits on $f$
with the $\sigma$ versus $f$ curves in \reffig{sigvf}, we obtain
the upper bounds on the magnification dispersion listed in
Table \ref{tbl:smax}.  

The bottom line is that microlensing is
basically negligible for image A, but reasonably important for
both images B and C.  For large source sizes, the central image C
is notably more affected by microlensing than the saddle image B,
hence microlensing of such demagnified images can help reveal the
source size.

\section{Discussion}

Central lensed images, which are highly demagnified and naturally
appear in places where the density of stars is high, are more
susceptible to gravitational microlensing than the more familiar
images that form at larger distances from lens galaxies.  Therefore,
it is useful to understand microlensing of central images in detail.

The microlensing magnification maps for demagnified images 
differ qualitatively from those for familiar magnified images in 
striking ways.  For high fractional surface density in stars, the 
central image maps do not show any of the classic fold and cusp 
caustic structures, but rather have concentrated blobs of
high magnification amid large regions of demagnification.  As $f$ is decreased 
the maps begin to develop more and more cusps and folds.  For the demagnified saddle, 
low values of $f$ yield cusps and folds, intermediate values produce blobs, and at very large 
values ($f \approx 100\%$) the cusps and folds reappear.  Thus the relative number of
cusps and folds versus blobs has a complicated dependence on $\kappa_{\rm tot}$, $\gamma$, and $f$ \citep[see also][]{PLW}. 

The central image C maps also show notable inhomogeneity on scales of tens of Einstein radii.
These large scale structures are evident in the 1$-$D auto-correlation function $\xi(\Delta x)$.  For the central image C, 
$\xi^{\rm C}$ falls to zero by $\Delta x_{0}^{\rm C} \approx 10.6 \Rein$ along the direction parallel to the shear, implying structure 
sizes $\sim 20$--$25 \Rein$.  In contrast $\Delta x_{0}^{\rm B} \approx 6.1$ suggesting smaller structures with 
sizes $\sim 10$--$15 \Rein$.
The larger scale structures in the image C maps 
causes microlensing fluctuations to be larger for
demagnified central images than for other images, even when the
source is fairly large ($\Rsrc/\Rein \ga 1$).

Our most intriguing qualitative result concerns the sensitivity of
central image microlensing to the relative densities of stars and
dark matter.  When the source is large ($\Rsrc/\Rein \ga 3$), the
magnification dispersion decreases monotonically with the fraction
$f$ of density in stars.  However, when the source is small
($\Rsrc/\Rein \la 3$), the magnification dispersion \emph{rises} as
$f$ is decreased from unity, reaches a peak at some finite value
of $f$, and then falls to zero as $f \to 0$ (as it should in the
absence of microlensing).  This dependence is similar to the behaviour
seen by \citet{SW} for a highly magnified saddle image.  We do not
see such behaviour for a demagnified saddle, but we do see it for a
demagnified central image.  While this qualitative result has little
practical importance (we expect $f \approx 1$ for central images),
it may be valuable as insight into how microlensing depends on the
relative amounts of stars and dark matter.  If we can develop a full
understanding of that problem, we may be able to turn microlensing
into the best tool available for probing local \emph{densities},
rather than integrated masses, of dark matter in distant galaxies
\citep[see][]{SW,SW04}.

One practical goal of microlensing studies is to probe the structure
of the optical continuum emission regions of quasars at very high effective
spatial resolution.  The light curves produced by relative motion
of the lens galaxy and source quasar can be used to map out the
(1-dimensional) structure of the quasar on micro-arcsecond scales
\citep{grieger,grieger2,agol,mineshige,fluke,GSGO}.  In terms of
raw variability amplitude, it would seem that central images are
the best targets for microlensing observations.  Of course, there
is the problem of detecting central images at optical wavelengths
in the first place.  A faint central image may be swamped by light
from the lens galaxy.  It may also be dimmed by extinction or
scattering in the interstellar medium of the lens galaxy, although
those effects might not be too much of a concern in the vast
majority of lens galaxies that are ellipticals.

It is interesting to consider the possibility that central image microlensing may allow us to probe the size of the \emph{radio} emission region of 
the source quasar.   For large sources the dispersion scales as  $\sigma \propto \Rsrc^{-1}$, so for radio sources with $\Rsrc/\Rein \ga 100$ we 
expect microlensing fluctuations to be no more than a few percent.\footnote{However, \citet{b1600} report that there is evidence for microlensing 
fluctuations in at least one radio lens, B1600$+$434.  Additionally, scintillation from the Milky Way may still have an appreciable effect at these 
source sizes for low galactic latitudes \citep[e.g.][]{koop-bigg}.}  For the case of PMN J1632$-$0033, the intensity of the central image is 
$\sim$750 and 500 $\mu$Jy at 8 and 22 GHz, respectively \citep{winn04}, so microlensing fluctuations may be at the level of tens of $\mu$Jy. 
Presently, integration times for observing these fluctuations are prohibitively long for a monitoring campaign ($\sim$5 or 50 hrs at 8 or 22 GHz to 
resolve 5\% fluctuations)\footnote{ c.f., http://www.vla.nrao.edu/astro/}. However, next generation telescopes such as the EVLA or EMERLIN will make 
these observations feasible.

Our results allow us to consider how microlensing may affect
various other applications that involve central lensed images.
In PMN J1632$-$0033, the position and brightness of the central
image lead to strong constraints on the density profile of the
lens galaxy, and to upper limits on the mass of any supermassive
black hole that may reside at the center of the lens galaxy
\citep{winn03,winn04}.  Those constraints are based on radio
data where, as noted above, the microlensing fluctuations are on the 
order of a few percent.  However, even if they were as large as
tens of percent, that would still be negligible compared with
the orders of magnitude over which the macro magnifications of central
images can vary due to modest changes in the smooth lens model
\citep[e.g.,][]{centers}.  In other words, microlensing does not
appear to be a significant concern for constraints on lens models
drawn from radio observations of central images.

Perhaps even more interesting is the possibility that any central
macro image produced by a lens galaxy containing a supermassive
black hole should be accompanied by a second, fainter central image
\citep{mao,bowman}, and that the central image pair can be used
to measure the black hole mass quite precisely \citep{rusin}.  The
second central image would be a saddle rather than a maximum, so
it would be interesting to consider whether a central saddle would
have different microlensing properties than the central maximum we
have considered.  We suspect that, at the radio wavelengths where
this application would be pursued, microlensing fluctuations are
still small compared with changes in the smooth lens model.
Nevertheless, it would be interesting to understand in detail how
microlensing of central saddles compares with microlensing of
central maxima.

\vspace{0.4cm}

\noindent
We are indebted to both anonymous referees for their helpful comments and suggestions.
We thank David Rusin and Josh Winn for stimulating our interest in
central images, and for providing results from lens models of PMN
J1632$-$0033.  We also thank Arthur Congdon for helpful discussions.
GGD and CRK are supported by grant HST-AR-10668
from the Space Telescope Science Institute, which is operated 
by the Association of Universities for Research in Astronomy,  
Inc., under NASA contract NAS5-26555.
JKW was supported by the European Community's Sixth Framework 
Marie Curie Research Training Network Programme, Contract No. 
MRTN-CT-2004-505183 "ANGLES".

\end{document}